\documentclass[10pt, twocolumn, fleqn]{article}

\usepackage[utf8]{inputenc}
\usepackage{amsmath, amssymb, mathtools, physics, bm}
\usepackage[low-sup]{subdepth}
\usepackage{enumitem}
\usepackage[toc]{appendix}
\usepackage{tensor}

\usepackage[letterpaper, left=0.75in, right=0.75in, top=1in, bottom=1in]{geometry}
\usepackage[labelfont=bf]{caption}
\usepackage{hyperref}  
\hypersetup{colorlinks, urlcolor=blue, citecolor=blue}
\usepackage{ftnright}
\usepackage{titlesec}
\titleformat{\section}{\large\bfseries}{\thesection.}{10pt}{\large}
\titleformat{\subsection}[runin]{\normalsize\bfseries}{\thesubsection}{8pt}{\normalsize}
\newcommand{\para}[1]{\subsection{#1.}}

\newcommand{\eq}[1]{\begin{equation} #1 \end{equation} }
\newcommand{\eqa}[1]{\begin{align*} #1 \numberthis \end{align*}}  
\newcommand{\eqn}[1]{\begin{equation*} #1 \end{equation*} }

\newcommand{\numberthis}{\addtocounter{equation}{1}\tag{\theequation}}

\newcommand{\La}{\mathcal{L}}
\newcommand{\Hu}{\mathcal{H}}
\newcommand{\p}{\partial}
\newcommand{\cov}{\nabla}

\newcommand{\A}{\mu}  
\newcommand{\B}{\nu}
\newcommand{\C}{\rho}
\newcommand{\D}{\sigma}
\newcommand{\E}{\alpha}
\newcommand{\F}{\beta}

\renewcommand{\expval}[1]{\langle #1 \rangle}

\DeclareMathOperator{\Si}{Si}
\DeclareMathOperator{\Ci}{Ci}

\usepackage{abstract}

\usepackage{color,soul}
\usepackage{cite}
\usepackage{ftnright}

\begin{document}

\twocolumn[{ 
\centering
\textbf{\Large Gauge-Invariant Scalar-Induced Gravitational Waves \\ from Physical Observables}

\vspace{15pt}
\large Vincent Comeau\footnotemark \vspace{5pt}

\textit{Department of Physics -- McGill University} \\
(\today)
\vspace{15pt}

\begin{abstract}
This paper discusses the gauge issue touching the gravitational waves induced at the second order by the scalar modes of cosmological perturbations. These waves are known to depend on the gauge used for their calculation. In this paper, we propose a simple method of obtaining physically meaningful expressions for such scalar-induced gravitational waves at the leading order. The method is centred on well-defined observables, such as the magnetic part of the Weyl tensor, or the Cotton tensor of a slicing of spacetime, which vanish in the background and do not depend linearly on the scalar perturbations. Generalizing the Stewart-Walker lemma, it is shown that the gravitational waves contributing to such observables at the second order are automatically gauge-invariant, even when the observable itself does not vanish at the first order. In each case, the scalar-induced gravitational waves are related to the ones computed in the Newtonian gauge, first for a general background, and then for the particular case of a spacetime dominated by either radiation or cold matter.
\vspace{12pt}
\end{abstract}
}]

\footnotetext{Email: \href{mailto:vincent.comeau@mail.mcgill.ca}{vincent.comeau@mail.mcgill.ca}}

{ \hypersetup{hidelinks} \tableofcontents }
\vspace{15pt}

\section{Introduction}

According to standard cosmology, our universe can be modelled on large scales as a nearly homogeneous expanding spacetime. At very early times, it contained small inhomogeneities, which can be classified into three types: the so-called scalar, vector, and tensor modes. These primordial perturbations evolve independently from one another when the Einstein equations are linearized, but mix with each other at higher orders, given the nonlinear nature of the Einstein equations. In particular, at the second order, the linear scalar perturbations become a source for the tensor perturbations.

Such scalar-induced gravitational waves have been extensively studied in recent years \cite{Domenech:2021, Baumann:2007, Kohri:2018}. Interest into such waves has been kindled after the experimental detection of gravitational waves by LIGO, in 2015. More recently, they have been considered as a potential source for the stochastic background of gravitational waves detected through pulsar timing arrays \cite{NANOGrav:2023a, NANOGrav:2023b, Antoniadis:2023, Reardon:2023, Xu:2023}. 

In almost all papers on this topic, the analysis is carried out along the following lines: first, the scalar-induced gravitational waves are computed by solving the relevant wave equation, obtained by expanding the Einstein equations up to the second order in a particular gauge. Their physical effect is then assessed by computing their energy density, or their power spectrum, using the standard formulas from the linear theory of gravitational waves.

However, as pointed out in \cite{Hwang:2017, Tomikawa:2019, Gong:2019, Gurian:2021}, such a procedure is flawed, as it yields different results when different gauges are used for the calculation. Indeed, tensor perturbations are invariant only at the first order in small coordinate transformations, but not at higher orders.

The purpose of this paper is to address the previous problem, by proposing a simple method of computing scalar-induced gravitational waves which are the same in all gauges, and bear a clear physical interpretation. Our method is centred on choosing physical observables which vanish in the homogeneous background, and only depend on the vector and tensor perturbations at the first order. For an observable of this kind, the terms induced at the second order by the linear scalar perturbations are gauge-invariant, as shown in section \ref{section-method} of this paper. These terms can thus be used to define scalar-induced gravitational waves which are physically meaningful, as they are closely connected by construction to a physical observable. 

In sections \ref{section-weyl} and \ref{section-cotton}, we apply our general method to two specific observables, namely the magnetic part of the Weyl tensor, and the Cotton tensor of a three-dimensional space-like surface. In particular, we show that the former observable provides a clear physical basis for the short-wavelength limit of the scalar-induced gravitational waves calculated in the Newtonian gauge, in the case of a spacetime dominated by radiation. This short-wavelength limit was quoted in \cite{NANOGrav:2023b}, when discussing the possible sources for the stochastic background of gravitational waves recently detected via a pulsar timing array.

However, we do not claim that the expressions we derive are gauge-invariant in the sense that the waves all share the same short-wavelength limit, whatever observable is used to define them. They are invariant, rather, \textit{for a given physical observable}. Ideally, one would hope to show that all the observables relevant to current detectors lead to waves which have the same short-wavelength limit, and are thus invariant in the first sense. We leave this important question to further work, and focus here on the invariance of the waves in the second sense.

\section{Review of Previous Work} \label{section-previous-work}

\para{First Approach} 

Various approaches have been proposed to deal with this gauge issue. First, some \cite{Chang:2020a, Chang:2020b} have constructed gauge-invariant quantities by combining the bare scalar-induced tensor perturbations $H_{ij}$ with additional terms designed to alleviate the gauge dependence, namely
\eq{
H_{ij}^\text{(GI)} = H_{ij} + \mathcal{X}_{ij}\,.
}

However, as argued in \cite{DeLuca:2019, Domenech:2020, Domenech:2021}, this approach does not solve the problem at all, since the quantities thus constructed are not unique, and do not bear any clear physical interpretation. In principle, for every such gauge-invariant quantity, there exists a particular gauge for which the counter-terms $\mathcal{X}_{ij}$ added to the bare tensor perturbations vanish, thus making the latter equal to the former. One can construct as many gauge-invariant quantities as there are possible gauges, and it is merely a matter of preference whether one chooses to work with the bare perturbation, rather than the gauge-invariant one.

The real issue at play here is not the gauge dependence of the scalar-induced gravitational waves, but their effect on a physical observable. Although all physical observables are necessarily gauge-invariant, as per the principle of relativity, not all gauge-invariant quantities are observables. Hence, in order to address the gauge issue affecting the scalar-induced gravitational waves, we must first and foremost consider observables, and only then ask what gauge-invariant perturbations can be constructed from them. This is the approach advocated in the present paper.

\para{Second Approach} 

On the other hand, some \cite{Inomata:2019, Yuan:2019, Domenech:2020, Ali:2020} have argued that the scalar-induced gravitational waves turn out to be the same in different gauges under certain circumstances, for instance in the short-wavelength limit, when the spacetime is dominated by radiation.

First, it must be acknowledged that the short-wavelength limit of the scalar-induced gravitational waves is not the same in all gauges, when all terms contributing to the waves are included. As shown in appendix \ref{appendix-short-wavelength}, the scalar-induced gravitational waves generated in a spacetime containing radiation exhibit radically different behaviours even in the short-wavelength limit, depending on the gauge used for their calculation. While they decay with $k\eta$ in the Newtonian and synchronous gauges, they oscillate without decaying in the isotropic comoving gauge, whereas they grow quadratically with $k\eta$ in the uniform-density gauge.

Nevertheless, in \cite{Domenech:2020}, it is shown that the scalar-induced gravitational waves share the same short-wavelength limit, when they are calculated in the subset of gauges which are ``well-behaved'' on small scales. A gauge is said to be ``well-behaved'' if the time slicing to which it corresponds on small scales is well-suited for the measurement of gravitational waves. The authors of \cite{Domenech:2020} then argue that the Newtonian gauge is well-behaved in this sense because, in that gauge and in the short-wavelength limit, Einstein's general relativity reduces to Newton's theory of gravity. It follows that a gauge is well-behaved if the metric perturbation $\phi$ and the fluid density perturbation are of the same order of magnitude as the corresponding quantities in the Newtonian gauge. 

For instance, these conditions are not satisfied for the uniform-density and the isotropic synchronous gauges considered in appendix \ref{appendix-short-wavelength}. Since the fluid density is not constant on small scales, the time slicing over which it is constant must be quite different from the one likely to be considered when measuring the gravitational waves. The divergent behaviour noticed in that gauge can thus be attributed to the behaviour of the slicing, rather than that of the gravitational waves themselves. 

The approach described in \cite{Domenech:2020} is quite convincing, as it is appropriately centred around the context of measurement, and the choice of a physical observable. However, this approach is not entirely satisfying as it is not based on any explicit observable, but only on more or less vague criteria regarding the time slicing chosen for the measurement. 

A similar approach has been followed in \cite{DeLuca:2019}, but with different conclusions. The authors of this paper argue that it is the synchronous gauge, rather than the Newtonian gauge, which is well-behaved with respect to the measurement of gravitational waves. Moreover, there certainly exist physical observables which might not be the ones measured by current detectors, but on which scalar-induced gravitational waves can have real effects, which cannot be ascribed to mere gauge artifacts. Hence, to avoid any confusion, it would be best to define clearly the observable we intend to measure, and only then consider the changes affecting this observable due to the scalar-induced gravitational waves.

Using a different argument, others have claimed \cite{Inomata:2019, Ali:2020} that the short-wavelength limit of the scalar-induced gravitational waves is the same in \textit{all} gauges, provided a formal distinction is made between tensor perturbations and gravitational waves. In their view, the latter name should be reserved for the components of the tensor perturbations which behave as free gravitational waves, generated in the absence of a source term from scalar perturbations.

For instance, the terms which dominate the short-wavelength limit of the perturbations in the isotropic comoving and uniform-density gauges, as computed in appendix \ref{appendix-short-wavelength}, should not be considered as proper gravitational waves according to \cite{Inomata:2019}, since they do not oscillate at the same frequency as the free waves, namely with frequencies $(\alpha\pm \beta)k\eta$ rather than $k\eta$. 

However, handpicking in this fashion the terms which make up the ``real'' gravitational waves seems entirely unjustified. At best, some gauges can be preferred to others, as in \cite{Domenech:2020, DeLuca:2019}, on the basis that they might be more relevant to observables; but, when a specific gauge is chosen for the calculation, there is no reason why some terms should be excluded in favour of others.

\para{Third Approach} \label{subsection-third-approach}

Finally, some \cite{Ota:2021} have proposed a more indirect approach, whereby the scalar-induced gravitational waves are assessed through their effects on a scalar observable, averaged over a physically meaningful surface. This surface could be one where a specific property of a fluid contained in the spacetime is held constant, for instance its density or its temperature. This is precisely the approach one follows when studying the back-reaction of cosmological perturbations on scalar observables, as done in \cite{Geshnizjani:2002, Kolb:2004, Comeau:2023}. 

In \cite{Ota:2021}, the scalar observable considered for the scalar-induced gravitational waves is the magnitude of the extrinsic curvature of a surface, averaged over that same surface. In particular, after taking the traceless and transverse component of the extrinsic curvature $K_{ij}$, one can construct a scalar observable which reduces to the usual energy density of gravitational waves at the first order,
\eq{
\rho_K =\, \expval{(K_{ij})^{\text{TT}} (K^{ij})^{\text{TT}}} \,\simeq\, \frac{1}{4a^2} \expval{\dot{H}_{ij}\,\dot{H}_{ij}}\,,
}
where a dot denotes a derivative with respect to the background conformal time. The magnitude of the extrinsic curvature does bear a clear physical interpretation, not only as a geometrical quantity, but also since it is involved in the Hamiltonian constraint relating the geometry of the surface to the energy density of matter. Hence, it might seem reasonable to interpret $\rho_K$ as the average energy density contained in gravitational waves, whether they be free or sourced by scalar perturbations. Of course, other scalar observables could also play that role, for instance the magnitude of the magnetic part of the Weyl tensor, which similarly reduces to the usual energy density of gravitational waves at the first order.

Although there is nothing wrong in principle with the approach we just described, it suffers from one main drawback. Namely, it is based on a global observable, defined in terms of an average over an entire slicing of spacetime. If possible, it would be much better to consider instead a purely local observable, involving no such average. However, as for most quantities, the extrinsic curvature of a surface is not itself gauge-invariant at the second order, according to the Stewart-Walker lemma, since its background value varies with time. Its magnitude becomes gauge-invariant only after it has been properly averaged. 

Nevertheless, there exists alternative quantities which vanish in the background, and from which can be constructed gauge-invariant expressions for the scalar-induced gravitational waves, thus making the averaging procedure superfluous. This is the approach advocated in the present paper.
 
\section{General Method} \label{section-method}

In this paper, we consider observables in a slightly inhomogeneous spacetime, which can be described as a spatially homogeneous background, onto which are added small perturbations. The spacetime metric can then be written as
\eq{
g_{\A\B} = g_{\A\B}^{(0)} + \delta g_{\A\B}\,,
}
where $g_{\A\B}^{(0)}$ denotes the metric of the background. Following the usual procedure, the perturbations can be decomposed into their so-called scalar, vector, and tensor components. As a result of the background homogeneity, these three types of perturbations decouple from one another at the first order: they are governed by separate parts of the linearized Einstein equations, and can thus be fixed by separate initial conditions.

When such a decoupling occurs at the first order, there exist certain observables which lead to gauge-invariant expressions for the second-order vector and tensor perturbations induced by the linear scalar perturbations. To show this, let us consider a general tensor $Q$, of any rank. This observable can be expanded in a series, in terms of its value $Q^{(0)}$ in the background, plus small perturbations,
\eq{
Q = Q^{(0)} + Q^{(1)} + Q^{(2)} + \cdots \,,
}
where $Q^{(1)}$ contains the terms linear in the metric perturbations, $Q^{(2)}$ those quadratic in the perturbations, and so on. 

Since the distinction between the background and the perturbations is not unique, the same coordinates in the background can be associated with two distinct points in the actual spacetime. In other words, if the coordinates used to label a point are changed from $x^\A$ to $x'^\A = x^\A + \epsilon^\A$, while keeping the same coordinates for the background, the value of the tensor $Q$ changes by the amount
\eq{
\delta_\epsilon Q = Q'(x) - Q(x)\,.
}
At the first order in the perturbations, we have
\eq{
\delta_\epsilon Q = - \La_\epsilon Q\,,
}
where $\La_\epsilon$ is the Lie derivative along the small displacement $\epsilon^\A$. Let us now consider the case of an observable which vanishes in the background, $Q^{(0)} = 0$, and whose perturbation at the first order only depends on the vector and tensor perturbations. Under these circumstances, the observable is gauge-invariant at the first order, since $\delta_\epsilon Q^{(1)} = - \La_\epsilon Q^{(0)} = 0$, although it ceases to be at higher orders. 

Moreover, its value at the first order is fixed by the initial conditions imposed on the vector and tensor metric perturbations, but not on those imposed on the scalar perturbations, since they evolve independently from one another. At the second order, the observable now involves corrections induced by all three types of perturbations, including a correction quadratic in the linear scalar perturbations. Hence, its value up to the second order can be decomposed as
\eqa{
Q &= \lambda_V \big(Q_V^{(1)} + Q_V^{(2)}\big) + \lambda_T \big(Q_T^{(1)} + Q_T^{(2)}\big) 
\\[3pt] &\hspace{30pt} + \lambda_S Q_S^{(2)}\,,
} 
where $Q_V$ and $Q_T$ denote the contributions from the linear vector and tensor perturbations, and $Q_S$ represents the correction induced at the second order by the linear scalar perturbations (which includes, in part, the scalar-induced vector and tensor perturbations). Here, $\lambda_S$, $\lambda_V$, and $\lambda_T$ are constants symbolizing the initial conditions imposed on each type of perturbations. Since the latter decouple at the first order, these constants can be fixed independently from another, which implies that the functions which they multiply transform independently under small coordinate transformations. At the second order, we thus get
\eqa{
&\delta_\epsilon Q_V^{(2)} = -\La_\epsilon Q_V^{(1)}\,, \\[3pt]
&\delta_\epsilon Q_T^{(2)} = -\La_\epsilon Q_T^{(1)}\,, \\[3pt]
&\delta_\epsilon Q_S^{(2)} = 0\,.
}

Hence, although the observable is not \textit{fully} gauge-invariant at the second order, the correction induced by the linear scalar perturbations is itself gauge-invariant at that order. This conclusion, which can be seen as a small generalization of the Stewart-Walker lemma \cite{Stewart:1974}, holds only as long as the observable vanishes in the background, and does not depend linearly on the scalar perturbations.

\section{Magnetic Part of the Weyl Tensor} \label{section-weyl} 

\para{Framework} 

In the rest of this paper, we consider two examples of physical observables satisfying the requirements outlined in the previous section, and thus involving gauge-invariant expressions for the scalar-induced vector and tensor perturbations.

Moreover, we consider a nearly homogeneous expanding spacetime, containing one or various perfect fluids. For simplicity, and given the current cosmological observations, the background is assumed to be spatially flat. The spacetime metric can then be written as
\eq{
g_{\A\B}(\eta, \bm{x}) = a^2(\eta)\big(\eta_{\A\B} + h_{\A\B}(\eta,\bm{x})\big)\,,
}
where $\eta$ is the conformal time of the background, $\bm{x}$ its spatial coordinates, $a$ its scale factor, and $\eta_{\A\B}$ is the Minkowski metric with $\eta_{00} = -1$. The background expansion rate is denoted by
\eq{
\Hu = \frac{\dot{a}}{a}\,,
}
where a dot represents a derivative with respect to the conformal time. Following the usual procedure, the metric perturbations can be decomposed in terms of their so-called scalar, vector, and tensor components,
\eqa{
h_{00} &= \phi \,,\\[3pt]
h_{0i} &= \p_i B + B_i \,,\\[3pt]
h_{ij} &= \delta_{ij}\psi + \p_i\p_j E + \p_i E_j + \p_j E_i + H_{ij}\,,
}
where $B_i$ and $E_i$ are the divergenceless vector perturbations, with $\p_i B_i = \p_i E_i = 0$, and $H_{ij}$ are the traceless and transverse tensor perturbations, with $H_{ii} = 0$ and $\p_i H_{ij} = 0$. Although the scalar and vector perturbations are not gauge-invariant, they can be combined with each other to form perturbations which are themselves invariant up to the first order in small coordinate transformations,
\eqa{
&\Psi = \psi + 2\Hu\sigma\,, \\[3pt]
&\Phi = \phi - 2\Hu\sigma - 2\dot{\sigma}\,, \\[3pt]
&\Phi_i = B_i - \dot{E}_i\,,
}
where $\sigma$ is defined as
\eq{
\sigma = B - \frac12 \dot{E}\,.
}

For a spacetime  of the kind considered in this paper, containing one or many perfect fluids, no anisotropic stress is generated at the first order, which implies that $\Psi = \Phi$. 

Furthermore, the upcoming calculation will involve various time-like vectors, for instance the velocities of the fluids contained in the spacetime. Let $v^\A$ be any such vector, which reduces in the background to the velocity of a comoving observer, and is normalized with $g_{\A\B} v^\A v^\B = -1$. Up to the first order, we then have
\eqa{ \label{velocity-first-order}
&v_0 = a\left(-1 + \frac12 \phi\right)\,,\\[4pt]
&v_i = a\left( \p_i v + V_i \right)\,,
}
where $V_i$ represents the vector part of the spatial velocity, with $\p_i V_i = 0$. The latter will play no role in the calculation, and can thus be neglected. On the other hand, the scalar part of the spatial velocity can be combined with the metric perturbations, to form a gauge-invariant  velocity potential,
\eq{ \label{velocity-invariant}
 V = v - \sigma\,.
}

The linearized Einstein equations constrain the spatial velocities of the various fluids contained in the spacetime only through their average, defined by
\eq{
u_i = a\left( \p_i u + U_i \right) = \frac{\sum\limits_A (\rho_A + p_A)\, u_{A\,i}}{\sum\limits_A (\rho_A + p_A)}\,,
}
where the sum is over the various fluids present in the spacetime, with $\rho_A$ and $p_A$ being their respective densities and pressures, and $u_{A\,i}$ their spatial velocities. The linearized Einstein equations then imply
\eq{ \label{ave-spatial-velocity}
U =\, u - \sigma \,=\, \frac{1}{2(\Hu^2 - \dot{\Hu})} (\dot{\Phi} + \Hu \Phi)\,.
}

Throughout this paper, we will often have to extract the traceless and transverse part of tensors, by acting on them with a projection operator \cite{Maggiore:2007}. Such an operator is well-defined in Fourier space, but we will also apply it to tensors in position space, in order to simplify the notation. By definition, it cancels any scalar or vector mode, namely $\Lambda_{ij}^{kl} \p_k = 0$ and $\Lambda_{ij}^{kl} \delta_{kl} = 0$. The traceless and transverse component of a tensor $T_{ij}$ is given by
\eq{
\Lambda_{ij}^{kl} T_{kl} (\eta, \bm{x}) = \int \frac{\dd[3]{k}}{(2\pi)^3} \,e^{i\bm{k}\cdot\bm{x}}\, \Lambda_{ij}^{kl} (\bm{k}) \,T_{kl}(\eta, \bm{k})\,. \qquad
}

\para{First-order Weyl tensor} 

The magnetic part of the Weyl tensor of the spacetime is a physical observable of the kind we discussed earlier, namely it vanishes in the background, and does not depend linearly on the scalar perturbations. 

The Weyl tensor at a point of spacetime corresponds to the traceless part of the Riemann tensor at that point. In four dimensions,
\eqa{
C_{\A\B\C\D} &= R_{\A\B\C\D} - \frac16 R \big(g_{\A\D} g_{\B\C} - g_{\A\C} g_{\B\D} \big) \\[2pt]
&+ \frac12 \big( g_{\A\D} R_{\B\C} + g_{\B\C} R_{\A\D} - g_{\A\C} R_{\B\D} - g_{\B\D} R_{\A\C}\big) \,,
}
where $R_{\A\B\C\D}$ is the Riemann tensor, $R_{\A\B} = \tensor{R}{^\C_\A_\C_\B}$ is the Ricci tensor, and $R=R_\A^\A$ the Ricci scalar. Moreover, the Weyl tensor is invariant under scale transformations, which makes it insensitive to the scale factor of the background. Writing the metric as $g_{\A\B} = a^2\,\hat{g}_{\A\B}$, with $\hat{g} = \eta_{\A\B} + h_{\A\B}$, we can show that
\eq{
\tensor{C}{^\A_\B_\C_\D} = \tensor{\hat{C}}{^\A_\B_\C_\D}\,.
}

As a result of its scale invariance, the Weyl tensor vanishes identically in the background. Moreover, by projecting its components along a time-like direction $v^\A$, it can be decomposed into its so-called ``electric'' and ``magnetic'' components, defined as \cite{Ellis:2012}
\eqa{
&E^{\A\B} = g^{\A\E} v^\C v^\D\, \tensor{C}{^\B_\C_\D_\E} \,, \\[4pt]
&B^{\A\B} = \eta^{\A\C\D\F} v_\F v^\E \tensor{C}{^\B_\E_\C_\D}\,,
}
where $\eta^{\A\B\C\D}$ denotes the Levi-Civita tensor\footnote{As explained in \cite{Ellis:2012}, the Levi-Civita tensor is related to the usual Levi-Civita symbol $\epsilon^{\A\B\C\D}$ in flat spacetime via
\eqn{
\eta^{\A\B\C\D} = \frac{1}{\sqrt{g}} \,\epsilon^{\A\B\C\D}\,.
}
}. This is analogous to the decomposition of the electromagnetic tensor in terms of an electric field and a magnetic field, hence the terminology employed. Here, we do not specify the vector $v^\A$, apart from asking that it be normalized, with $g_{\A\B} v^\A v^\B = -1$. It could represent, for instance, the velocity of one of the fluids contained in the spacetime.

Furthermore, the scalar-induced gravitational waves contribute to the Weyl tensor, at the second order, only through terms which are symmetric in their indices\footnote{Indeed, at the second order, the linear scalar perturbations contribute to the antisymmetric part of any observable via terms of the form
\eqn{
A_{ij} = f \,\p_i \p_j g - g \,\p_i \p_j f\,,
}
where $f$ and $g$ are two functions, of the same order of magnitude as the linear scalar perturbations. These terms do not contain any transverse and traceless component, which could contribute to the scalar-induced gravitational waves,
\eqn{
A_{ij}^\text{TT} = \Lambda_{ij}^{kl} A_{kl} = 0\,.
} 
}. Hence, for the purposes of the present paper, it suffices to take the symmetric part of the magnetic part of the Weyl tensor, which we redefine as
\eq{ \label{def-magnetic-weyl}
B^{\A\B} = \eta^{\A\C\D\F} v_\F v^\E \tensor{C}{^\B_\E_\C_\D} \,+\, (\A\leftrightarrow\B)\,.
}

At the first order, the electric part of the Weyl tensor depends on all three types of metric perturbations, whereas its magnetic part only depends on the vector and tensor perturbations,
\eqa{
&B^{00} = 0\,, \\[3pt]
&B^{0i} = 0\,, \\[3pt]
&B^{ij} = \frac{1}{a^4} \,\epsilon_{ikl} \,\p_k ( \dot{H}_{jl} - \p_j \Phi_l ) \,+\, (i\leftrightarrow j)\,,
}
where $\epsilon_{ijk}$ is the usual Levi-Civita symbol in three-dimensional flat space. As the magnetic part of the Weyl tensor does not depend linearly on the scalar perturbations, we expect the terms generated at the second order by these perturbations to be gauge-invariant. 

The scalar-induced gravitational waves thus obtained bear a clear physical interpretation, as they are directly related to an observable expressing the curvature of spacetime. Moreover, the spatial average of the magnitude of this observable coincides, at the leading order, with the standard formula for the energy density contained in the various Fourier modes of the waves,
\eq{
\rho_B = \expval{B_{\A\B} B^{\A\B}} \,\simeq \,-\frac{4}{a^4}\expval{\dot{H}_{ij}\cov^2 \dot{H}_{ij}} \,.
}

\para{Gauge-invariant perturbation} \label{subsection-weyl-invariant}

Let us now compute the contribution to the magnetic part of the Weyl tensor induced by the scalar perturbations. To do so, we expand (\ref{def-magnetic-weyl}) up to the second order, keeping all terms quadratic in the scalar perturbations, while neglecting the ones quadratic in the vector and tensor perturbations, as well as the ones mixing the perturbations with one another. The scalar-induced magnetic part of the Weyl tensor is then given by
\eqa{
&B^{00} = 0\,, \\[3pt]
&B^{0i} = 0\,, \\[3pt]
&B^{ij} = \frac{1}{a^4} \,\epsilon_{ikl} \,\p_k ( \dot{H}_{jl}^B - \p_j \Phi_l^B ) \,+\, (i\leftrightarrow j)\,,
}
where $\Phi_i^B$ and $H_{ij}^B$ represent the second-order scalar-induced vector and tensor perturbations related to the magnetic part of the Weyl tensor. In particular, the tensor component takes the form
\eqa{ \label{wave-magnetic-weyl-1}
\dot{H}_{ij}^B &= \dot{H}_{ij} + \Lambda_{ij}^{kl}\Big( 2V\,\p_k\p_l \Phi - 2\sigma\,\p_k\p_l \dot{\sigma} \\[3pt] 
&\hspace{20pt} - \dot{\psi}\,\p_k\p_l E  - \psi\,\p_k\p_l \dot{E} + \frac12 \p_n E \,\p_k\p_l\p_n \dot{E} \Big)\,.
}

By construction, this expression is gauge-invariant up to the second order in small coordinate transformations. Another gauge-invariant expression for these perturbations, albeit not related to a physical observable as in this case, has been discussed in \cite{Chang:2020a, Domenech:2020}. When written properly, it is given by
\eqa{ \label{wave-newtonian}
H_{ij}^N &= H_{ij} - \Lambda_{ij}^{kl}\Big( \sigma \,\p_k\p_l \sigma + \psi\,\p_k \p_l E \\[3pt]
&\hspace{70pt} - \frac14 \p_n E \,\p_k\p_l \p_n E \Big)\,.
}

These perturbations correspond to the scalar-induced gravitational waves $H_{ij}$ calculated in the Newtonian gauge, for which $\sigma=E=0$, hence the notation. For some reason, the expression we provide here seems to diverge from the ones given in \cite{Chang:2020a, Domenech:2020}, which also do not agree with each other so far as we can tell. We assume mere typos or differences in notation are responsible for this discrepancy. As shown in appendix \ref{appendix-gauge-invariance}, our expression is indeed correct, as it is gauge-invariant up to the second order in small coordinate transformations. Moreover, in a spacetime containing various perfect fluids, these perturbations satisfy the wave equation
\eqa{ \label{wave-equation-newtonian}
&\ddot{H}_{ij}^N + 2\Hu\,\dot{H}_{ij}^N - \cov^2 H_{ij}^N \\[3pt]
&\hspace{12pt} = -\Lambda_{ij}^{kl}\Big( \Phi\,\p_k\p_l\Phi + 4(\Hu^2 - \dot{\Hu})\,U\,\p_k\p_l U \Big)\,,
}
where $U$ is the scalar component of the average spatial velocity of the fluids, as given in (\ref{ave-spatial-velocity}). Combining (\ref{velocity-invariant}), (\ref{wave-magnetic-weyl-1}), and (\ref{wave-newtonian}), we get
\eq{ 
\dot{H}_{ij}^B = \dot{H}_{ij}^N + 2\,\Lambda_{ij}^{kl}\big( V\,\p_k\p_l \Phi \big)\,.
} 

Since $H_{ij}^N$ is known to be gauge-invariant up to the second order, the same is then true of $H_{ij}^B$, which confirms the general analysis carried out in section \ref{section-method}. Integrating with respect to time, we get
\eq{ \label{wave-magnetic-weyl}
H_{ij}^B = H_{ij}^N + 2 \,\Lambda_{ij}^{kl} \int \dd{\eta}\, V\,\p_k\p_l \Phi + C_{ij}\,,
}
where $C_{ij}$ is the integration constant, which can be fixed using the initial conditions imposed on the perturbations. However, as argued in \cite{Domenech:2020}, the physical observables relevant for gravitational waves are expected to be time-dependent, and thus insensitive to constant terms such as $C_{ij}$. Indeed, both the energy density of gravitational waves, and the magnetic part of the Weyl tensor, involve $\dot{H}_{ij}$ rather than $H_{ij}$, which annihilates any constant term.

\para{Choice of initial conditions} \label{subsection-initial-conditions}

Let us now compute the scalar-induced gravitational waves related to the magnetic part of the Weyl tensor, for the particular case of a spacetime dominated either by cold matter or radiation. To do so, we could solve the wave equation (\ref{wave-equation-newtonian}) satisfied by the tensor perturbations in the Newtonian gauge, using for instance the appropriate Green's function. Here, we will use instead the well-known solutions obtained in \cite{Kohri:2018}.

When computing the scalar-induced gravitational waves generated during a certain stage of the history of the universe, either when it is dominated by radiation or cold matter, we will neglect any waves which might have been generated in previous stages. More specifically, in the simplified case of a spacetime containing a single fluid, we choose to set $H_{ij} = 0$ at the beginning of the expansion, when $\eta = 0$. This is the convention adopted in \cite{Kohri:2018}, as in most papers on the topic. 

Furthermore, it is convenient to write the scalar-induced gravitational waves in terms of their Fourier modes, with wavenumber $\bm{k}$, as
\eqa{ \label{def-I}
H_{ij} (\eta, \bm{k}) &= \int \frac{\dd[3]{q}}{(2\pi)^3} \,\Phi_{\bm{q}}(0)\,\Phi_{\bm{k} - \bm{q}}(0) \\[3pt] &\hspace{30pt}\times \,\frac{\Lambda_{ij}^{kl}(\bm{k})\,q_k q_l}{k^2} \, I(\eta, \bm{k}, \bm{q})\,,
}
where $\Phi_{\bm{k}}(0)$ represents the initial value of the scalar metric perturbation, and $I$ is a function expressing the evolution of the waves with time. Similarly, we have
\eqa{
\big(V\,\p_i \p_j \Phi\big)(\eta, \bm{k}) &= -\frac12 \int \frac{\dd[3]{q}}{(2\pi)^3} \,\Phi_{\bm{q}}(0)\,\Phi_{\bm{k} - \bm{q}}(0) \\[3pt]
&\hspace{25pt} \times \, q_i q_j \,\Big( \hat{V}_{\bm{q}}\,\hat{\Phi}_{\bm{k-q}} + \hat{V}_{\bm{k-q}}\,\hat{\Phi}_{\bm{q}}\Big)\,,
}
where a hat, when applied to a perturbation $f$, indicates that the initial value of the scalar metric perturbation has been factored out from it,
\eqa{ \label{def-f-hat}
&\hat{f}_{\bm{k}}(\eta) = \frac{f_{\bm{k}}(\eta)}{\Phi_{\bm{k}}(0)}\,.
}

Plugging these expressions in (\ref{wave-magnetic-weyl}), the scalar-induced gravitational waves related to the magnetic part of the Weyl tensor become
\eq{ \label{I-magnetic-weyl}
I^B = I^N - k^2 \int_0^\eta \dd{\eta} \Big( \hat{V}_{\bm{q}}\,\hat{\Phi}_{\bm{k-q}} + \hat{V}_{\bm{k-q}}\,\hat{\Phi}_{\bm{q}}\Big)\,.
}

\para{Application to cold matter}

Let us now consider the case when the Weyl tensor is decomposed into its magnetic and electric parts, by being projected along the velocity of the fluid contained in the spacetime, $V=U$. The latter is computed using (\ref{ave-spatial-velocity}).

When the spacetime contains cold matter ($w=0$), the scalar metric perturbation is constant,
\eqa{
&\hat{\Phi} = 1\,, \\[3pt]
&\hat{U} = \frac16 \eta \,,
}
which implies that
\eq{
I^B = I^N - \frac16 x^2\,,
}
where $x=k\eta$. The scalar-induced gravitational waves have been computed in the Newtonian gauge in \cite{Kohri:2018}. Adjusting the overall multiplicative constant of their equation (37) in order to fit our conventions, we get
\eq{ \label{newtonian-cold-matter}
I^N = \frac53 + \frac{5}{x^2} \cos x - \frac{5}{x^3} \sin x \,,
}
which implies that
\eq{
I^B = - \frac16 x^2 + \frac53 + \frac{5}{x^2}\cos x - \frac{5}{x^3}\sin x\,.
}

Hence, contrary to the Newtonian scalar-induced gravitational waves, which remain finite in the short-wavelength limit, the ones related to the magnetic part of the Weyl tensor do not, but instead grow linearly with the scale factor. This kind of behaviour has been noticed before, but typically dismissed as being a mere gauge artefact, not representative of a real gravitational wave on the basis that it does not involve any oscillation \cite{Ali:2020}. 

Here, however, the growing mode is clearly not a result of the gauge used for the calculation; it has real, measurable effects on the observable to which it is related by construction. The only way of possibly getting rid of this growing mode would be to project the Weyl tensor along a different physically meaningful direction, for instance along the velocity of a clock field contained in the spacetime, but having a negligible effect on its expansion rate.

\para{Application to radiation}

Similarly, when the spacetime contains radiation ($w=\frac13$), plugging (\ref{ave-spatial-velocity}) in (\ref{I-magnetic-weyl}) yields
\eq{ \label{I-magnetic-radiation}
I^B = I^N - \frac14 x^2 \,\hat{\Phi}_{\bm{q}}\,\hat{\Phi}_{\bm{k}-\bm{q}}\,,
}
where $x=k\eta$. In the short-wavelength limit, when $x$ is large, the scalar-induced gravitational waves in the Newtonian gauge decay as $\frac{1}{x}$, whereas the scalar metric perturbation decays as $\frac{1}{x^2}$. Hence, although the waves related to the magnetic part of the Weyl tensor do not coincide exactly with the ones calculated in the Newtonian gauge, they reduce to the latter in the short-wavelength limit: $I^B \simeq I^N$ when $x$ is large.

The short-wavelength limit calculated in the Newtonian gauge can truly be considered as physical, and not merely gauge-dependent, since it also applies to the gravitational waves contributing to the magnetic part of the Weyl tensor, which is an observable.

More specifically, the scalar-induced gravitational waves in the Newtonian gauge are given by equation (22) of \cite{Kohri:2018}. Combining this solution with (\ref{I-magnetic-radiation}), we get
\eqa{
\hspace{-15pt}I^B &= \frac{9(\alpha^2 + \beta^2 - 1)}{8\alpha^3\beta^3\, x^2}\Big(\sin(\alpha x) \sin(\beta x) - \alpha\beta x \sin x\Big) \\[3pt]
&+\frac{9(\alpha^2 + \beta^2 - 1)^2}{32\alpha^3\beta^3\, x} \Big( f\big((1 + \alpha + \beta)x\big) + f\big((1 - \alpha - \beta)x\big) \\[3pt]
&\hspace{30pt} - f\big((1 + \alpha - \beta)x\big) - f\big((1 - \alpha + \beta)x\big) \Big)\,,
}
where $f$ is defined as
\eq{
f(x) = \Si(x) \cos x + \Big(\ln(\abs{x}) - \Ci(\abs{x})\Big) \sin x\,,
}
and $\alpha$ and $\beta$ are dimensionless quantities, defined as\footnote{These quantities differ only by a factor of $\sqrt{3}$ from the variables $u$ and $v$ typically defined in the literature, as in \cite{Kohri:2018}. We did not want to use $u$ and $v$, in order to avoid any confusion with the velocities also denoted by these symbols.}
\eqa{ \label{alpha-beta}
\alpha &= \frac{q}{\sqrt{3}\,k}\,, \\[3pt]
\beta &= \frac{\abs{\bm{k} - \bm{q}}}{\sqrt{3}\,k}\,.
}

In the long-wavelength limit, when $x$ is small, the waves are given by
\eq{
I^B = -\frac{9(\alpha^2 + \beta^2 - 1)^2}{8\alpha^2\beta^2} \,x^2\,,
}
whereas, in the short-wavelength limit, they become\footnote{To obtain this short-wavelength limit, we must pay attention to the sign of the argument of the sine integral functions, knowing that $\alpha$ and $\beta$ are both positive, and that $\abs{\alpha-\beta}<1$. This inequality can be derived by applying the triangle inequality to the vectors $\bm{k}$, $\bm{q}$, and $\bm{k} - \bm{q}$.}
\eqa{
\hspace{-5pt}I^B &= - \frac{9\pi(\alpha^2 + \beta^2 - 1)^2}{32\alpha^3\beta^3}\,\Theta(\alpha + \beta - 1) \,\frac{\cos x}{x} \\[3pt]
&+ \frac{9(\alpha^2 + \beta^2 - 1)^2}{32\alpha^3\beta^3}\, \ln\left|\frac{1-(\alpha+\beta)^2}{1-(\alpha-\beta)^2}\right|\,\frac{\sin x}{x} \\[3pt]
&-\frac{9(\alpha^2 + \beta^2 - 1)}{8\alpha^2\beta^2}\,\frac{\sin x}{x} \,,
}
where $\Theta$ denotes the usual Heaviside step function. As we argued earlier, this coincides with the  short-wavelength limit of the scalar-induced gravitational waves computed in the Newtonian gauge, and often encountered in the literature.

\section{Cotton Tensor} \label{section-cotton}

\para{Induced metric} 

The Cotton tensor of a slicing of spacetime represents another observable of the same kind as the magnetic part of the Weyl tensor, which also leads to a gauge-invariant expression for the scalar-induced gravitational waves. 

In this case, however, the Cotton tensor is not defined relative to the full four-dimensional spacetime, as the Weyl tensor, but with respect to a certain three-dimensional space-like surface. For this surface to be physically meaningful, its definition must involve some property of the fluids contained in the spacetime, for instance the density or the temperature of one of their components. 

To keep our calculation as general as possible, we simply assume that some scalar property $\chi$ of the fluids is held constant over the surface. This fluid property can be written in terms of its value $\chi_0$ in the background, plus a small perturbation,
\eq{
\chi(\eta, \bm{x}) = \chi_0(\eta) + \delta\chi(\eta, \bm{x})\,.
}

We can also construct a time-like vector $v^\A$, normalized with $g_{\A\B} v^\A v^\B = -1$, which is everywhere normal to the surface,
\eq{
v_\A = - \frac{\p_\A \chi}{\sqrt{-g^{\B\C}\, \p_\B \chi\, \p_\C \chi}}\,.
}

Expanding up to the first order, we get the same expressions as in (\ref{velocity-first-order}), but with the scalar component of the spatial velocity given by
\eq{ \label{scalar-spatial-chi}
v = -\frac{\delta\chi}{\dot{\chi}_0}\,.
}

Moreover, since the surface is defined in terms of a single scalar function, there exists certain coordinates for which it coincides perfectly with the surfaces of constant background time, namely when $v=0$. More generally, when working in an arbitrary gauge, we can always change the background time coordinate from $\eta$ to $\eta' = \eta - v$, while keeping the same spatial coordinates, in order to make the surfaces of constant $\eta'$ and $\chi$ coincide with each other. The latter are then described by a metric $g_{ij}'$, defined as
\eq{
g_{ij}'(\eta', \bm{x}) = a^2(\eta')\big( \delta_{ij} + h_{ij}'\big)\,.
}

The perturbations affecting the induced metric of the surface can be related to the original metric perturbations, expressed in an arbitrary gauge, via the standard transformation rule for tensors (\ref{transformation-rule-metric}). In particular, when changing the background conformal time to $\eta' = \eta - v$, the scalar components of the perturbations become, up to the first order\footnote{These formulas can be obtained directly from the equations given in appendix \ref{appendix-gauge-invariance}, that is, by taking $\alpha = v$ and $\beta = 0$ in (\ref{linear-delta-epsilon}) and (\ref{quadratic-delta-epsilon}).},
\eqa{ \label{scalar-new-time}
&\psi' = \psi + 2\Hu\,v\,, \\[3pt]
&E' = E\,,
}
whereas the tensor perturbations become, up to the second order,
\eq{ \label{tensor-new-time}
H_{ij}' = H_{ij} + \Lambda_{ij}^{kl}\Big( (2\Hu E - 2\sigma + v)\,\p_k\p_l v \Big)\,.
}

\para{Gauge-invariant perturbation}

The surface we just described possesses measurable geometrical properties, pertaining for instance to its intrinsic curvature. Since the surface has three dimensions, its Weyl tensor vanishes identically, and so its Riemann tensor can be expressed in terms of its Ricci tensor. 

Moreover, although the latter vanishes in the background, it is not a suitable observable leading to gauge-invariant scalar-induced gravitational waves, as it depends linearly on the scalar perturbations. Hence, the simplest observable suitable for such a purpose must involve derivatives of the Ricci tensor. This is the case of the Cotton tensor, whose properties are discussed for instance in \cite{York:1971}. Keeping only its symmetric part, as we did for the magnetic part of the Weyl tensor, it is defined as
\eq{
C^{ij} = \eta^{ikl}\,\cov_k' \tensor{{R'}}{_l^j} \,+\, (i\leftrightarrow j) \,,
}
where $\cov_i'$ represents the covariant derivative relative to the surface with an induced metric $g_{ij}'$, and $R_{ij}'$ is its Ricci tensor. Expanding up to the first order, we get
\eq{
C^{ij} = -\frac{1}{2a^5}\,\epsilon_{ikl}\,\p_k \cov^2 H_{jl} \,+\, (i\leftrightarrow j) \,.
}

Hence, as the Cotton tensor depends only on the tensor perturbations at the first order, we expect the contribution induced at the second order by the scalar perturbations to be gauge-invariant. Expanding up to that order, we get indeed
\eq{
C^{ij} = -\frac{1}{2a^5}\,\epsilon_{ikl}\,\p_k \cov^2 H_{jl}^C \,+\, (i\leftrightarrow j) \,,
}
where $H_{ij}^C$ represents the scalar-induced gravitational waves contributing to the Cotton tensor,
\eq{
H_{ij}^C = H_{ij}' \,+\, \Lambda_{ij}^{kl}\,\Big( -\psi'\,\p_k\p_l E' \,+\, \frac14 \p_n E'\,\p_k\p_l\p_n E'\Big)\,,
}
that is, using (\ref{velocity-invariant}), (\ref{scalar-new-time}), and (\ref{tensor-new-time}),
\eq{
H_{ij}^C = H_{ij}^N + \Lambda_{ij}^{kl}\,\big( V\,\p_k\p_l V\big)\,.
}

As expected, the scalar-induced gravitational waves related to the Cotton tensor of a slicing are really gauge-invariant at the second order. They reduce to the bare tensor perturbations in the specific gauge for which $v=0$, \textit{and} $E=0$. Contrary to what one might expect, using coordinates for which the surfaces of constant background time and fluid property $\chi$ coincide with each other is necessary, but not sufficient to insure that $H_{ij}$ coincides with the gauge-invariant expression we just obtained.

Taking the Fourier transform of the previous equation, at a wavenumber $\bm{k}$, and factoring out the initial value of the scalar metric perturbation as in (\ref{def-I}), we then get 
\eq{ \label{I-cotton}
I^C = I^N - k^2 \,\hat{V}_{\bm{q}} \hat{V}_{\bm{k}-\bm{q}}\,.
}

\para{Application to cold matter}

As a simple example, let us consider the case of a spacetime containing a single fluid, whose density serves as the property held constant over the surface on which the Cotton tensor is computed. As given in (\ref{scalar-spatial-chi}), the spatial component of the vector perpendicular to the surface of constant fluid density is then
\eq{
v = -\frac{\delta\rho}{\dot{\rho}_0}\,,
}
where $\rho_0$ is the background fluid density, and $\delta\rho$ its perturbation. Using (\ref{velocity-invariant}), as well as the Friedmann equations for the background, and the linearized Einstein equations, we get
\eq{
V = \frac{1}{2(\Hu^2-\dot{\Hu})}\,\Big(\dot{\Phi} + \Hu\Phi - \frac{1}{3\Hu}\,\cov^2\Phi \Big)\,.
}

For instance, when the spacetime contains cold matter ($w=0$), the scalar metric perturbation is constant, which implies that
\eq{
\hat{V} = \frac16\,\eta\,\Big(1 + \frac{1}{12} x^2\Big)\,.
}

Combining this equation with (\ref{newtonian-cold-matter}) and (\ref{I-cotton}), the scalar-induced gravitational waves related to the Cotton tensor for the surface of constant density are then given by
\eqa{
I^C &= -\frac{1}{5184}\,x^6 - \frac{1}{216}\,x^4 - \frac{1}{36}\,x^2 + \frac53 \\[4pt]
&\hspace{53pt} + \frac{5}{x^2}\cos x - \frac{5}{x^3}\sin x\,.
}

Once again, the growing terms included in the waves are not mere gauge artefacts, since they would be obtained whatever gauge were used for the calculation. Of course, they might be attributable to the surface for which we chose to compute the Cotton tensor. Such a surface might not be suited for the measurement of gravitational waves by current detectors, since the density of matter is not constant on small scales, as argued in \cite{Domenech:2020}. However, this does not change the fact that the growing modes we just obtained are truly physical, and have effects on an observable which could be measured in principle.

\section{Conclusion}

In this paper, we have shown how physically meaningful, gauge-invariant expressions for the scalar-induced gravitational waves can be constructed from a certain type of observables. More specifically, following our small generalization of the Stewart-Walker lemma, the scalar-induced gravitational waves contributing to an observable are gauge-invariant if the latter vanishes in the background, and does not depend linearly on the scalar perturbations. 

In particular, for the case of a spacetime dominated by radiation, we showed that the waves related to the magnetic part of the Weyl tensor have the same short-wavelength limit as the ones computed in the Newtonian gauge. So far as we are aware, this result provides the first explicit connection between the short-wavelength limit obtained in the Newtonian gauge, and an actual observable. Such a physical basis is important, as this limit has been frequently featured in the literature, for instance in \cite{NANOGrav:2023b} regarding the stochastic background of gravitational waves recently detected using a pulsar timing array.

More generally, in previous work on this topic, the gauge issue affecting the scalar-induced gravitational waves has often been discussed without any reference to observables. At best, it has been addressed by describing the kind of gauges which might correspond to observables suitable for the measurement of gravitational waves, but which are never explicitly defined. In our view, it would be best to define the observable first, and only then proceed with the calculation.

Of course, the two observables considered in this paper --- the magnetic part of the Weyl tensor, and the Cotton tensor of a slicing --- are merely examples of quantities which lead straightforwardly to gauge-invariant scalar-induced gravitational waves, without having to take a spatial average. Other observables could be considered, which might be more relevant to the measurement of gravitational waves by current detectors. We welcome any proposals on that matter.

Finally, it should be noted that our method cannot be applied to the gravitational waves induced at the second order by the mixing of the linear scalar and tensor perturbations, as recently discussed in \cite{Bari:2023}. In this case, the induced perturbations could be characterized by following the third approach described in section \ref{subsection-third-approach} of this paper, that is, by considering their effects on the average value of a scalar observable.

\newpage
\section*{Acknowledgments}

The work done for this paper was supported in part by the Fonds de recherche du Qu\'ebec (FRQNT). I have also benefited greatly from discussions with my PhD advisor Robert Brandenberger (McGill University). Moreover, I would like to thank Misao Sasaki and Antonio Riotto for their insightful comments on a previous version of this paper.
\vspace{20pt}

\begin{appendices}
\section{Short-Wavelength Behaviours} \label{appendix-short-wavelength}

In this appendix, we compute the scalar-induced gravitational waves produced in a spacetime dominated by radiation, using three different gauges. Our goal is to confirm that the short-wavelength limit of these waves can be radically different, depending on the gauge used for their calculation, as argued in \cite{Tomikawa:2019, Gong:2019, Gurian:2021}. 

Although this represents a well-known conclusion, our calculation appears simpler than previous ones, as it is based on a gauge-invariant quantity. The scalar-induced gravitational waves in any gauge can be deduced from their expression in the Newtonian gauge via (\ref{wave-newtonian}),
\eqa{
H_{ij}^N &= H_{ij} - \Lambda_{ij}^{kl}\Big( \sigma \,\p_k\p_l \sigma + \psi \,\p_k \p_l E \\[3pt]
&\hspace{70pt} - \frac14 \p_n E \,\p_k\p_l \p_n E \Big)\,.
}

Following our notation in (\ref{def-I}) and (\ref{def-f-hat}), we can rewrite the Fourier modes of the waves in terms of
\eqa{
I &= I^N - k^2 \Big( \hat{\sigma}_{\bm{q}} \hat{\sigma}_{\bm{k} - \bm{q}} + \frac12 \hat{\psi}_{\bm{q}} \hat{E}_{\bm{k} - \bm{q}} + \frac12 \hat{\psi}_{\bm{k} - \bm{q}} \hat{E}_{\bm{q}} 
\\[3pt] &\hspace{70pt} + \frac14 (\bm{k}\cdot\bm{q}-q^2)\hat{E}_{\bm{q}} \hat{E}_{\bm{k} - \bm{q}} \Big)
}

Let's use this equation to compute the short-wavelength limit of the scalar-induced gravitational waves in three different gauges, for a spacetime containing radiation ($w=\frac13$). In doing so, we will systematically neglect any terms decaying faster than the waves in the Newtonian gauge, namely faster than $\frac{1}{x}$, where $x=k\eta$. 

\para{Isotropic comoving gauge} 

In the isotropic comoving gauge, for which $v=0$ and $E=0$, the relevant scalar perturbation is given by
\eq{
\hat{\sigma}_{\bm{q}} = \frac{3}{4k\alpha} \Big( \sin(\alpha x) + \frac{2}{\alpha x} \cos(\alpha x) - \frac{2}{\alpha^2 x^2} \sin(\alpha x) \Big)\,,
}
and similarly for $\hat{\sigma}_{\bm{k} - \bm{q}}$, but with $\beta$ instead of $\alpha$. The latter are the dimensionless quantities defined in (\ref{alpha-beta}). Hence, in the short-wavelength limit, when $x\gg1$, the scalar-induced gravitational waves in that gauge oscillate without decaying, and thus they do not reduce to the ones computed in the Newtonian gauge,
\eqa{
I &= -\frac{9}{16\alpha\beta} \sin(\alpha x)\sin(\beta x) -\frac{9}{8\alpha^2\beta\, x} \cos(\alpha x) \sin(\beta x) \\[3pt] &\hspace{20pt} - \frac{9}{8\alpha\beta^2\, x} \sin(\alpha x) \cos(\beta x) + I_N \,.
}

\para{Uniform-density gauge}

In the uniform-density gauge, for which $\delta\rho = 0$ and $E=0$, the density of the radiation is fixed to its background value. In that case, the relevant scalar perturbation is
\eqa{
\hat{\sigma}_{\bm{q}} &= \frac34\, \eta\, \Big( \cos(\alpha x) - \frac{2}{\alpha x} \sin(\alpha x) 
\\[3pt]&\hspace{30pt} - \frac{2}{\alpha^2 x^2} \cos(\alpha x) + \frac{2}{\alpha^3 x^3} \sin(\alpha x) \Big)\,,
}
which implies, in the short-wavelength limit,
\eqa{
I &= -\frac{9}{16}\,x^2 \cos(\alpha x)\cos(\beta x) + \frac{9}{8\alpha}\,x\sin(\alpha x)\cos(\beta x) \\[3pt]
&\hspace{10pt}+ \frac{9}{8\beta}\,x\cos(\alpha x)\sin(\beta x) - \frac{9}{4\alpha\beta}\,\sin(\alpha x)\sin(\beta x) \\[3pt]
&\hspace{10pt}+ \frac{9(\alpha^2 + \beta^2)}{8\alpha^2\beta^2}\,\cos(\alpha x)\cos(\beta x) 
\\[3pt] 
&\hspace{10pt}- \frac{9(\alpha^2 + 2\beta^2)}{8\alpha^2\beta^3\,x}\,\cos(\alpha x)\sin(\beta x) 
\\[3pt]
&\hspace{10pt}- \frac{9(\beta^2 + 2\alpha^2)}{8\alpha^3\beta^2\,x}\,\sin(\alpha x)\cos(\beta x)\,+\,I_N \,.
}

Hence, the scalar-induced gravitational waves in that gauge grow quadratically with the scale factor, rather than decaying as the ones in the Newtonian gauge. 

\para{Synchronous gauge}

Finally, in the synchronous gauge, for which $\phi = 0$ and $B=0$, the relevant scalar perturbations are
\eqa{
&\hspace{-5pt}\hat{\sigma}_{\bm{q}} = -\frac{3}{2k\alpha^2\, x}\left( C - \frac{\sin(\alpha x)}{\alpha x}\right)\,, \\[3pt]
&\hspace{-5pt}\hat{E}_{\bm{q}} = \frac{3}{k^2\alpha^2}\left( C\,\ln(\alpha x) + \frac{\sin(\alpha x)}{\alpha x} - \Ci(\alpha x)\right)\,,
}
where $C$ is an integration constant, which is not fixed by the constraints imposed in the synchronous gauge, and thus represents a residual gauge degree of freedom. In principle, the value chosen for this constant should not matter, as it should not have any effect on physical observables. 

However, insofar as computing bare scalar-induced gravitational waves makes sense, we can fix this constant to $C=0$, by requiring that the waves be ``well-behaved'' in the short-wavelength limit, as argued in \cite{Domenech:2020}. We can also take $C=1$, by imposing the same initial conditions on the waves as we did in section \ref{subsection-initial-conditions}, namely that they remain small initially, at $\eta=0$. In the short-wavelength limit, we get
\eq{
I = \frac{9(3\alpha^2+3\beta^2-1)}{8\alpha^2\beta^2} \,C^2\,\ln(\alpha x)\,\ln(\beta x) + I_N\,.
}

Hence, the scalar-induced gravitational waves computed in the synchronous gauge grow logarithmically with the scale factor if $C=1$, whereas they reduce to the ones in the Newtonian gauge if $C=0$, but then they do not satisfy the same initial conditions as the latter.

\section{Proof of Gauge Invariance} \label{appendix-gauge-invariance}

In this appendix, we show that the ``Newtonian'' scalar-induced tensor perturbations, discussed in section \ref{subsection-weyl-invariant} of this paper, are indeed gauge-invariant up to the second order in small coordinate transformations,
\eqa{ \label{newtonian-invariant-appendix}
H_{ij}^N &= H_{ij} - \Lambda_{ij}^{kl}\Big( \sigma \,\p_k\p_l \sigma + \psi \,\p_k \p_l E \\[3pt]
&\hspace{70pt} - \frac14 \p_n E \,\p_k\p_l \p_n E \Big)\,.
}

To do so, we consider a small coordinate transformation $x'^\A = x^\A + \epsilon^\A$, where $\epsilon^\A$ is of the same order of magnitude as the metric perturbations. Moreover, we are essentially concerned here with the gauge invariance of the perturbations with respect to the scalar degrees of freedom. The divergenceless part of the transformation affecting the spatial coordinates can thus be set to zero, as it amounts to choosing a gauge for the linear vector perturbations. Following the notation in \cite{Bruni:1996}, we consider the transformation
\eqa{
\epsilon^0 &= -\alpha\,, \\[3pt]
\epsilon^i &= -\p_i \beta\,.
}

Under a coordinate transformation, the components of the metric at a point of spacetime become
\eq{ \label{transformation-rule-metric}
g_{\A\B}'(x') = \pdv{x^\C}{x'^\A}\,\pdv{x^\D}{x'^\B}\, g_{\C\D}(x)\,.
}
Expanding in a series, we get
\eq{
\delta_\epsilon g_{\A\B} = g_{\A\B}'(x) - g_{\A\B}(x)\,.
}

In particular, the scalar components of the metric perturbations change by small amounts given, at the first order, by the well-known formulas
\eqa{ \label{linear-delta-epsilon}
&\delta_\epsilon \phi = -2\dot{\alpha} - 2\Hu\alpha \,, \\[3pt]
&\delta_\epsilon \psi = 2\Hu\alpha\,, \\[3pt]
&\delta_\epsilon \sigma = -\alpha\,, \\[3pt]
&\delta_\epsilon E = 2\beta\,.
}

The tensor components of the metric perturbations are invariant at the first order, but not at higher orders. The amount by which they change at the second order has been computed in \cite{Bruni:1996}. Taking the tensor component of their equation (5.37), and adapting the notation, we get
\eqa{ \label{quadratic-delta-epsilon}
\delta_\epsilon H_{ij} &= \Lambda_{ij}^{kl}\Big( (-2\sigma + \alpha + 2\Hu E + 4\Hu\beta)\,\p_k\p_l \alpha \\[2pt]
&\hspace{10pt} + 2\psi\,\p_k\p_l \beta - \p_n (E + \beta)\,\p_k\p_l\p_n \beta\Big) \,.
}

Combining this equation with (\ref{newtonian-invariant-appendix}) and (\ref{linear-delta-epsilon}), we find that the ``Newtonian'' tensor perturbations are indeed independent of the gauge chosen for the scalar perturbations, $\delta_\epsilon H_{ij}^N = 0$.

\end{appendices}

\bibliographystyle{mybibstyle}
\bibliography{references_induced_GWs}

\end{document}